\documentclass{iau_FM}
\usepackage{graphicx}
\usepackage{graphicx}
\usepackage{natbib}
\usepackage{amsmath}
\usepackage{amssymb}
\usepackage{arydshln}
\usepackage{txfonts} 
\usepackage{caption}

\DeclareRobustCommand{\ion}[2]{%
	\relax\ifmmode
	\ifx\testbx\f@series
	{\mathbf{#1\,\mathsc{#2}}}\else
	{\mathrm{#1\,\mathsc{#2}}}\fi
	\else\textup{#1\,{\mdseries\textsc{#2}}}%
	\fi}

\title[Stellar feedback at low metallicity
] 
{Massive star feedback in the Magellanic Clouds and the tidal Bridge}

\author[Varsha Ramachandran
 et al]   
{Varsha Ramachandran
$^1$}

\affiliation{$^1$Zentrum f{\"u}r Astronomie der Universit{\"a}t Heidelberg,
Astronomisches Rechen-Institut, M{\"o}nchhofstr. 12-14, 69120 Heidelberg \\ [\affilskip]}

\pubyear{2022}
\jname{Astronomy in Focus, Focus Meeting 4} 
\editors{José Espinosa, ed}
\begin{document}

\maketitle

\begin{abstract}
Massive stars have far-reaching feedback effects that alter the surrounding environment on local, global, and cosmic scales. Spectral analyses of massive stars with adequate stellar-atmosphere models are important to study massive star feedback in detail. We discuss the most recent UV and optical studies of massive metal-poor stars, including those with metallicities ranging from half to one twentieth of solar, connected with large-scale ISM structures in the Magellanic Clouds and the tidal Magellanic Bridge. We present ionizing fluxes from massive stars with low metallicity along with mechanical energy, and we further compare these to the observed energetics in the ISM. The results give hints on the leakage of hot gas and ionizing photons in the Magellanic Clouds. The paper outlines feedback from individual massive stars to population-level collective feedback, the significance of various feedback mechanisms (radiation, wind, supernova), and the influence  by the physical conditions of the ISM.
\end{abstract}

\firstsection 
\section{Introduction}

Stellar feedback is one of the largest uncertainties in star and galaxy formation. Massive stars are of great interest because of their far-reaching feedback effects that alter the surrounding environment on local, global, and cosmic scales. The combined feedback in massive young stellar clusters leads to the
formation of superbubbles that can drive galactic winds and outflows, which have been frequently observed in local as well as distant galaxies.  The Magellanic Clouds and the Bridge offer an outstanding opportunity to investigate low metallicity massive stars and study feedback under conditions typical for the vast majority of dwarf galaxies. 

  \vspace{-0.3cm}
\section{Massive star feedback in the Magellanic Clouds}

We carried out spectroscopic observations of  $\sim\!500$ OB stars in the Magellanic Clouds using VLT-FLAMES. When available, the optical spectroscopy was complemented by UV spectra from the HST, IUE, and FUSE archives.  The two representative  young stellar populations that have been studied are associated with the superbubble N\,206 in the Large Magellanic Cloud (LMC)  and with the supergiant shell SMC-SGS\,1 in the Wing of the Small Magellanic Cloud (SMC), respectively. We performed  spectroscopic analyses of the massive stars using the  non-LTE Potsdam Wolf-Rayet (PoWR) model atmosphere code. We estimated the stellar, wind, and feedback parameters of the individual massive stars.
 
The total energy feedback from OB stars, WR stars and  supernovae in the  N\,206 and SGS complexes are compared in Table\,1. Since the winds of OB stars in the LMC are much stronger than in the SMC,  wind feedback in N\,206 is dominated by massive Of stars \citep{Ramachandran2018,Ramachandran2018b}. The stellar wind feedback from the two WR stars in N\,206 is comparable to that from all young OB stars together. The situation is completely different in the SMC supergiant shell, where the ionizing and mechanical luminosity is dominated by one WO star \citep{Ramachandran2019}.  The observed H$\alpha$ emission reflects almost half of the ionizing flux provided by the massive stars -- the other half of the LyC photons obviously escapes. The X-ray superbubble in the LMC is equally powered by stellar winds and supernovae. In contrast, the overall energy feedback of the supergiant shell in the SMC is dominated by supernovae, implying that hundreds of OB stars do not have much impact on the feedback before their final explosion. Thus, at low metallicities,  feedback is chiefly governed by supernova explosions and only few very massive stars.  The comparison of the total stellar input with X-ray, radio and H$\alpha$ observations shows that only a fraction the input energy accumulated over time is currently  still present in these regions.  The rest might have escaped or leaked out of the complex. Our result has significant importance for feedback in  low-metallicity galaxies, where we can now neglect the contribution from OB stars. However, at LMC like metallicities, young OB stars have a significant contribution in terms of ionizing and mechanical feedback. In that case, neglecting their contribution and considering only supernovae is not justified.  Concluding, the metallicity  decides whether  the stellar winds or supernovae become the key agents of feedback.

\begin{minipage}{\textwidth}
\begin{minipage}[b]{0.48\textwidth}
    \centering
    
   \begin{tabular}{crr}
\hline
\hline
\noalign{\vspace{1mm}}
Total $ E_{\mathrm{ mec}}$  &  {N206-LMC}&{SMC-SGS\,1} \\
\noalign{\vspace{1mm}}
 \hline 
 \noalign{\vspace{1mm}}
OB   &  $ 10^{52}$ (43\%)  &  $6 \times 10^{49}$ (0.5\%)  \\
WR  &  $ 10^{51}$ (9\%)  &  $2 \times 10^{51}$ (16\%)  \\
SNe  &  $1.2 \times 10^{52}$ (48\%)  &  $10^{52}$ (83\%)  \\
\noalign{\vspace{1mm}}
\hdashline
\noalign{\vspace{1mm}}
Observed & $4.7 \times 10^{51}$ (20\%)  &  $6 \times 10^{51}$ (50\%)  \\
\noalign{\vspace{1mm}}
\hline
\end{tabular} 
 \captionof{table}{Total mechanical energy feedback}
    \end{minipage}
  \begin{minipage}[b]{0.51\textwidth}
    \centering
    \includegraphics[scale=0.35, trim={0.2cm 1.7cm 0cm  0cm}]{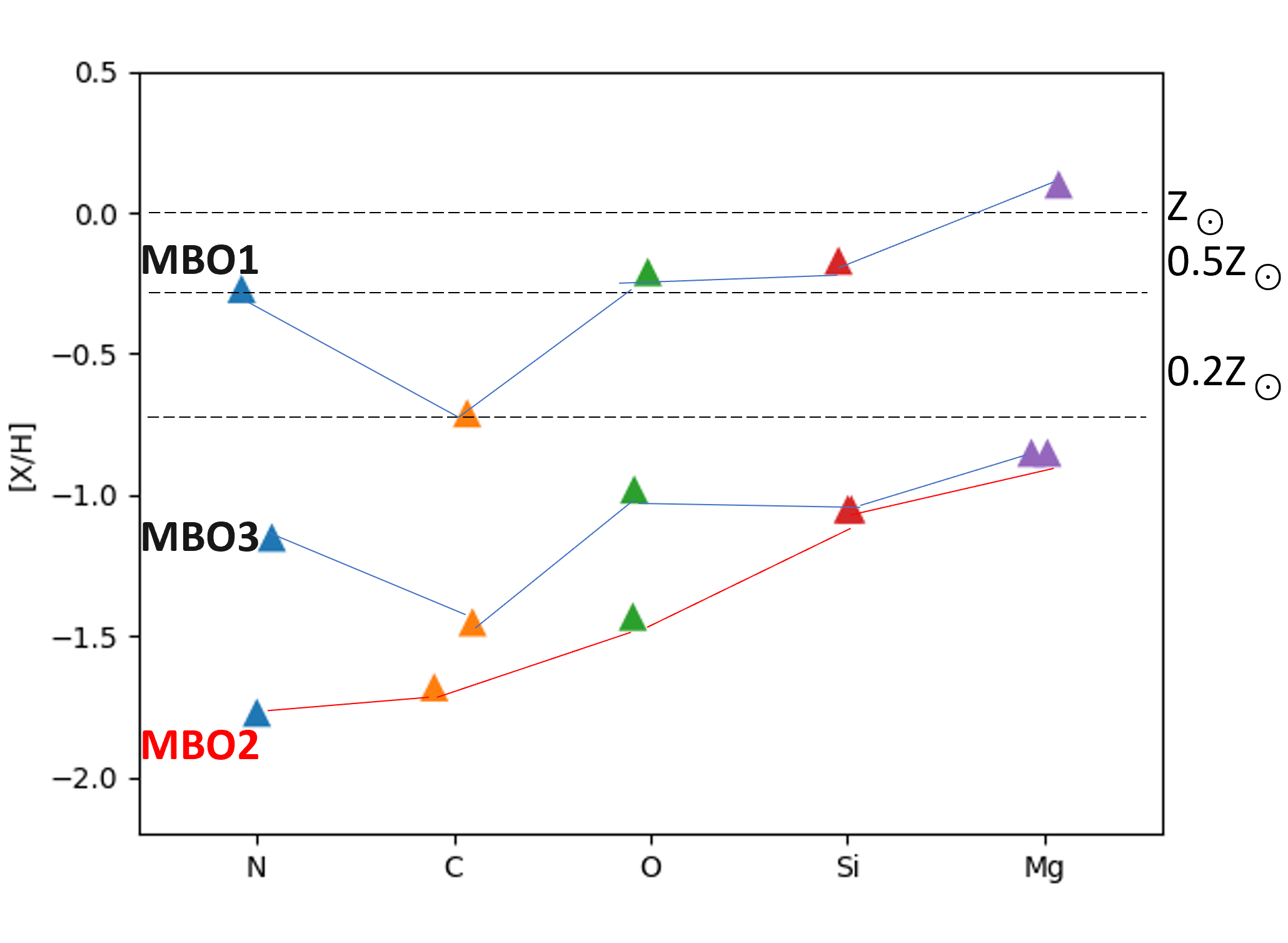} 
    \captionof{figure}{Chemical abundance in Bridge O stars}
  \end{minipage}
  \hfill
  
  \end{minipage}
  \vspace{-0.3cm}
\section{Metallicity and ionization of the Bridge}

The Magellanic Bridge, stretching between SMC and LMC,  is the nearest tidally stripped intergalactic environment. The Bridge has a significantly low average metallicity than SMC. For the first time we discovered three massive O stars in the Bridge using VL/FLAMES spectra \citep{Ramachandran2021}. We analyze the spectra of each star using the PoWR models, providing stellar parameters, ionizing photon fluxes, and surface abundances. The ages of the newly discovered O stars suggest that star formation in the Bridge is ongoing. The multi-epoch spectra indicate that all three O stars are binaries. Despite their spatial proximity to one another, these O stars are chemically distinct. MBO1 is a fast-rotating giant with nearly LMC-like abundances. The other two are main-sequence stars that rotate extremely slowly and are strongly metal depleted (Fig\,1). Among this MBO2 the most nitrogen-poor O star known to date. Taking into account the previous analyses of B stars in the Bridge, we interpret the various metal abundances as the signature of a chemically inhomogeneous ISM, suggesting that the Bridge gas might have accreted during multiple episodes of tidal interaction between the Clouds. Attributing the lowest derived metal content to the primordial gas, the time of the initial formation of the Bridge may date back several billion years. Using the Gaia and Galex color-magnitude diagrams, we roughly estimate the total number of O stars in the Bridge and their total ionizing radiation. Comparing this with the energetics of the diffuse ISM, we find that the contribution of the hot stars to the ionizing radiation field in the Bridge is less than 10\% and conclude that the main sources of ionizing photons are leaks from the LMC and SMC. This provide a lower limit for the fraction of ionizing radiation that escapes from these two dwarf galaxies.

\vspace{-0.5cm}
\bibliographystyle{aa} 
\bibliography{ref} 

\begin{thebibliography}{4}
\expandafter\ifx\csname natexlab\endcsname\relax\def\natexlab#1{#1}\fi

\bibitem[{{Ramachandran} {et~al.}(2018{\natexlab{a}}){Ramachandran}, {Hainich},
  {Hamann}, {Oskinova}, {Shenar}, {Sander}, {Todt}, \&
  {Gallagher}}]{Ramachandran2018}
{Ramachandran}, V., {Hainich}, R., {Hamann}, W.-R., {et~al.}
  2018{\natexlab{a}}, \aap, 609, A7

\bibitem[{{Ramachandran} {et~al.}(2018{\natexlab{b}}){Ramachandran}, {Hamann},
  {Hainich}, {Oskinova}, {Shenar}, {Sander}, {Todt}, \&
  {Gallagher}}]{Ramachandran2018b}
{Ramachandran}, V., {Hamann}, W.~R., {Hainich}, R., {et~al.}
  2018{\natexlab{b}}, \aap, 615, A40

\bibitem[{Ramachandran {et~al.}(2019)Ramachandran, Hamann, Oskinova, Gallagher,
  Hainich, Shenar, Sand~er, Todt, \& Fulmer}]{Ramachandran2019}
Ramachandran, V., Hamann, W.~R., Oskinova, L.~M., {et~al.} 2019, A\&A, 625,
  A104

\bibitem[{{Ramachandran} {et~al.}(2021){Ramachandran}, {Oskinova}, \&
  {Hamann}}]{Ramachandran2021}
{Ramachandran}, V., {Oskinova}, L.~M., \& {Hamann}, W.~R. 2021, \aap, 646, A16

\end{thebibliography}
\end{document}